\documentclass[11pt]{article}

\usepackage{amssymb,amsfonts,amsmath}

\usepackage{pstricks,pst-node,pst-text,pst-3d}

\newtheorem{theorem}{Theorem}[section]
\newtheorem{lemma}[theorem]{Lemma}
\newtheorem{proposition}[theorem]{Proposition}
\newtheorem{example}[theorem]{Example}
\newtheorem{corollary}[theorem]{Corollary}
\newtheorem{remark}[theorem]{Remark}

\newcommand{\bp}{\noindent {\em Proof. }}
\newcommand{\ep}{\hfill $\Box$ \par\medskip}

\def\calA{{\mathcal A}}

\def\calF{{\mathcal F}}
\def\calG{{\mathcal G}}

\def\calK{{\mathcal K}}
\def\calL{{\mathcal L}}

\def\calN{{\mathcal N}}
\def\calO{{\mathcal O}}

\def\bfA{{\mathbf A}}
\def\bfB{{\mathbf B}}

\def\bfI{{\mathbf I}}

\def\bfd{{\mathbf d}}

\def\bff{{\mathbf f}}

\def\bfk{{\mathbf k}}

\def\bfm{{\mathbf m}}
\def\bfn{{\mathbf n}}

\def\bfz{{\mathbf z}}

\def\bf0{{\mathbf 0}}
\def\bf1{{\mathbf 1}}
\def\bf2{{\mathbf 2}}
\def\bf3{{\mathbf 3}}
\def\bf4{{\mathbf 4}}
\def\bf5{{\mathbf 5}}
\def\bf6{{\mathbf 6}}
\def\bf7{{\mathbf 7}}
\def\bf8{{\mathbf 8}}
\def\bf9{{\mathbf 9}}

\def\ra{\rightarrow}

\def\LRa{\Leftrightarrow}

\def\es{\emptyset}
\def\se{\subseteq}

\def\ve{\varepsilon}

\def\vp{\varphi}
\def\vpi{\varphi^{-1}}

\def\DD{\mathbf{DD}}
\def\nDD{\mathbf{nDD}}
\def\LDD{\mathcal{D\!D}}
\def\nLDD{n\mathcal{D\!D}}
\def\DW{\mathcal{DW}}

\def\lg{\mathrm{lg}}
\def\ran{\mathrm{ran}}

\def\supp{\mathrm{supp}}

\def\faA{\mathcal{A} = (A,X,\delta)}
\def\frA{\bfA = (A,X,\delta,a_0,F)}
\def\An{\calA^{\bfn\bfd}}

\def\ndN{\mathcal{N} = (A,X,\alpha)}

\def\fzF{\mathcal{F} = (A,X,f)}
\def\fzG{\mathcal{G} = (B,X,g)}

\def\Fnd{\calF^{\bfn\bfd}}
\def\Fnda{\Fnd = (A,X,\alpha)}

\def\Afz{\calA^{\bff\bfz}}

\def\Nfz{\calN^{\bff\bfz}}
\def\Nfza{\Nfz = (A,X,f)}

\def\Dir{\mathbf{Dir}}

\def\flK{\calK = \{\calK(X)\}_X}
\def\flL{\calL = \{\calL(X)\}_X}

\begin{document}

\title{Directing Words of Fuzzy Finite Automata }

\author{Magnus Steinby\\
Department of Mathematics and Statistics\\
University of Turku\\
FI-20014 Turku, Finland}
\date{\today}

\maketitle

\begin{abstract}
A deterministic finite automaton is  \emph{directable} if it has a \emph{directing word} which takes the automaton from every state to the same  state. These notions have been extended also to other kinds of automata. Thus, B.~Imreh and M.~Steinby (1999) identified three natural types of directing words, called D1--, D2-- and D3--directing words, for nondeterministic finite automata (NFAs). Here we adapt these notions for fuzzy finite automata (FFAs). The D3--directing words obtained this way are precisely 
the directing words introduced by V.~Karthikeyan and M.~Rajasekar (2015). With any FFA $\calF$ we associate an NFA $\Fnd$ which has the same D$i$--directing words as $\calF$.  Thus, if these definitions are used, the theory of directable FFAs reduces to that of NFAs.

We also introduce three new kinds of directing words of fuzzy automata that we call DD1--, DD2-- and DD3--directing words, respectively which depend more on the fuzzy transition degrees between states.   We establish some basic properties of the sets $DD_i(\calF)$ of DD$i$--directing words of any given FFA $\calF$. In particular, it is shown that these languages are regular, and that DD$i$--directability is decidable. For so-called \emph{normal FFA}s the languages $DD_i(\calF)$ are shown to have some special properties. Several relationships between the families of the corresponding sets $DD_i(\calF)$ of DD$i$--directing words are presented. We also determine the complete $\cap$-semilattice of  the various classes of DD$i$--directable FFAs and normal FFAs and their intersections.

\end{abstract}

\smallskip



\section{Introduction}\label{Introduction}

An input word $w$ of a deterministic finite automaton (DFA) $\calA$ is said to be \emph{directing}, or \emph{synchronizing}, if it takes $\calA$ from every state to a fixed state $c$, and $\calA$ is \emph{directable} if it has directing words. The topic has been extensively studied ever since it was introduced by \v{C}ern\'{y} \cite{Cer64} in 1964. Directability has also been defined for other kinds of automata. In particular,  Imreh and Steinby \cite{ImSt99} identify three types of directing words of nondeterministic finite automata (NFAs).
In \cite{KaRa15} \text{Karthikeyan} and \text{Rajasekar} define  directing words for fuzzy finite automata (FFAs), and using this notion they attempt to fuzzify the contents of \cite{ImSt95} (without mentioning  this paper).

The three types of directing words of NFAs, the D1--, D2-- and D3--directing words, considered in \cite{ImSt99}, can be defined in a natural way for fuzzy automata, too. Then the directing words of \cite{KaRa15} are the same as the D3--directing words.
We show that for every FFA there is an NFA such that the two automata have exactly the same directability properties.  Hence, all the results aimed at in \cite{KaRa15}, and much more, follow from known facts about  NFAs (cf. \cite{Bur76,DoZa17,GaIvNG09,GoHeKoRy82,ImIS03,ImSt99,Ito04,ItST04,Sha18}, for example).\smallskip

Then we introduce three new types of directing words of fuzzy automata that  depend more on the transition degrees between states. The study of the sets of such directing words, the families of languages formed by them, and the classes of the corresponding directable FFAs form the main contents of this paper.

Section 2 introduces some basic terminology, notation and the automata to be considered.
In Section 3 we first recall the definitions of directing words for DFAs as well as the D1--, D2-- and D3--directing words of NFAs. Then we define D1--, D2-- and D3--directing words of FFAs.  With each FFA $\calF$ we associate an NFA $\Fnd$ that faithfully reflects some important properties of the original automaton. In particular, for each $i = 1,2,3$, the FFA $\calF$ and the NFA $\Fnd$ have the same D$i$--directing words, which means that the properties of D$i$--directable FFAs and their directing words can be derived directly from the theory of D$i$--directable NFAs. We also note that for any FFA $\calF$, the languages $D_1(\calF)$, $D_2(\calF)$ and $D_3(\calF)$ of D1--, D2-- and D3--directing words of $\calF$ are regular, and that D$i$--directability is therefore decidable. Finally, it is shown how an algorithm for testing the directability of a DFA presented in \cite{ImSt95} can be modified for deciding whether a given (complete) FFA is D3--directable.

In Section 4 we introduce DD1--, DD2-- and DD3--directing words of FFAs, and DD1--, DD2-- and DD3--directable FFAs, and present some properties of the sets $DD_1(\calF)$, $DD_2(\calF)$ and $DD_3(\calF)$ of the respective sets of directing words of an FFA $\calF$. Among other matters, we exhibit relationships between the languages $DD_i(\calF)$ and $D_i(\calF)$. Also the languages $DD_i(\calF)$ are shown to be regular.

In Section 5, we study the relationships between the families of languages formed by the sets $DD_i(\calF)$ of DD$i$-directing words of such FFAs. In Section 6 we establish the inclusion relationships between the classes of DD$i$-directable FFAs and the classes of
DD$i$-directable so-called normal FFAs. \smallskip

Some familiarity with the basic theory of
finite automata and regular languages is assumed (cf. \cite{Arb69}, \cite{HoMoUl07},  \cite{Koz97}, \cite{Sak09} or \cite{Sal69}, for example).  For the general theory
of fuzzy automata and fuzzy languages, the reader may consult \cite{MoMa02} or \cite{Rah09}. The papers \cite{BoImCiPe99} and \cite{Vol08} survey parts of the theory of directable automata. Much of our terminology and notation stems from \cite{ImSt95,ImSt99}.


\section{Some basic notions}\label{Basic}

We may write $A := B$ to emphasize that $A$ is defined to be equal to $B$. The cardinality of a set $A$ is denoted by $|A|$, and the set of all subsets of $A$ by $\wp(A)$. For any integer $n \geq 0$, let $\bfn := \{1,\ldots,n\}$.


In what follows, $X$ is always a finite nonempty alphabet. The set of all (finite) words over $X$ is denoted by $X^*$ and the empty word by $\ve$. Furthermore, let $X^+ := X^*\setminus \{\ve\}$. The length of a word $w \in X^*$ is denoted by $\lg(w)$. Subsets of $X^*$ are called \emph{languages}. A \emph{family of languages} $\calL$ assigns to each finite nonempty alphabet $X$ a set $\calL(X)$ of languages over $X$. We write $\calL = \{\calL(X)\}_X$.

A \emph{fuzzy language} over $X$ is a mapping $\lambda : X^* \ra [0,1]$, where the set of \emph{degrees of membership} is the real unit interval $[0,1]$ ordered by the usual $\leq$-relation and equipped with the lattice operations $r \vee s = \max(r,s)$ and $r\wedge s = \min(r,s)$. Thus we follow Zadeh's \cite{Zad65} original definition.
The \emph{support} of $\lambda$, is the language $\supp(\lambda) := \{w \in X^* \mid \lambda(w) > 0\}$, and its  \emph{range} is the set $\ran(\lambda) := \{\lambda(w) \mid w \in X^*\}$.  If $\supp(\lambda)$ is a finite set $\{w_1,\ldots,w_n\}$, we may write\vspace{-3pt}
\[
\lambda = \{w_1/\lambda(w_1),\ldots,w_n/\lambda(w_n)\}.\vspace{-3pt}
\]
If $\ran(\lambda) \se \{0,1\}$, then $\lambda$ is said to be \emph{crisp}.


A \emph{deterministic finite automaton}, a \emph{DFA} for short, $\faA$ consists of a finite nonempty set $A$ of \emph{states}, the \emph{input alphabet} $X$, and a \emph{transition function} $\delta : A\times X \ra A$ which  is extended in the usual way to a mapping $\delta^* : A\times X^* \ra A$.

A \emph{deterministic finite recognizer (DFR)} $\frA$ is a DFA $\faA$ equipped with an \emph{initial state} $a_0\in A$ and a set $F\se A$ of \emph{final states}. It \emph{recognizes} the language $L(\bfA) := \{w\in X^* \mid \delta^*(a_0,w) \in F\}$, and a language $L\se X^*$ is said to be \emph{recognizable}, or \emph{regular}, if $L = L(\bfA)$ for such a recognizer $\bfA$. The set of regular languages over $X$ is denoted by $Rec(X)$.\smallskip

A \emph{nondeterministic finite automaton (NFA)} $\ndN$ consists of  a finite nonempty set $A$ of \emph{states}, the \emph{input alphabet} $X$, and  a nondeterministic \emph{transition function} $\alpha : A \times X \ra \wp(A)$. The transition function is extended to a mapping
$
\alpha^* : \wp(A) \times X^* \ra \wp(A)
$
as follows: for every $H\se A$,
\begin{itemize}
  \item[(1)] $\alpha^*(H,\ve) = H$,  and
  \item[(2)] $\alpha^*(H,ux) = \bigcup\{\alpha(b,x) \mid b \in \alpha^*(H,u)\}$ for any  $u \in X^*$ and $x\in X$.
\end{itemize}
For any $H \se A$ and $w\in X^*$, $\alpha^*(H,w)$ is the set
of states reachable by the input word $w$ from some state in $H$. For any $a\in A$ and $w\in X^*$, we write $\alpha^*(a,w)$ for $\alpha^*(\{a\},w)$.
The NFA $\calN$ is {\it complete}, if
$\alpha(a,x) \not =\emptyset $ for all $a\in A$ and $x\in X$. Obviously, then $\alpha^*(a,w) \not = \es$ for all $a\in A$ and $w\in X^*$.\smallskip

A \emph{fuzzy finite automaton (FFA)} $\fzF$ consists of a finite nonempty set $A$ of \emph{states}, the \emph{input alphabet} $X$, and  a fuzzy \emph{transition function} \vspace{-3pt}
\[
f : A\times X \times A \ra [0,1]. \vspace{-3pt}
\]
In concrete examples of an FFA $\fzF$, we specify the transition function $f$ by giving the nonzero values $f(a,x,b)$.

The transition function is extended to  $f^* : A\times X^* \times A \ra [0,1]$ by setting
\begin{itemize} \vspace{-3pt}
  \item[(1)] $f^*(a,\ve,a) = 1$, and $f^*(a,\ve,b) = 0$ if $b\neq a$, and  \vspace{-3pt}
  \item[(2)] $f^*(a,w,b) = \max\{f^*(a,v,c) \land f(c,x,b)\}\mid c\in A\}$
  for $w = vx$ with $v\in X^*$ and $x\in X$,  \vspace{-3pt}
\end{itemize}
for all $a,b\in A$.  For any $a\in A$ and $w\in X^*$, let \vspace{-3pt}
\[
\calF(a,w) := \{b\in A \mid f^*(a,w,b) > 0\}. \vspace{-3pt}
\]
Thus $\calF(a,w)$ is the set of states reachable (to a nonzero degree) from state $a$ by the input word $w$.
With any pair $a\in A$ and $w \in X^*$, we also associate a fuzzy subset \vspace{-3pt}
\[
\calF_{a,w} : A \ra [0,1], \: b \mapsto f^*(a,w,b), \vspace{-3pt}
\]
of $A$.  Obviously, $\supp(\calF_{a,w}) = \calF(a,w)$.

For any $H \se A$ and $w \in X^*$, let $\calF(H,w) := \bigcup\{\calF(a,w) \mid a \in H\}$.

The fuzzy automaton $\calF$ is \emph{complete} if $\calF(a,x) \neq \es$ for all $a\in A$ and $x\in X$. If $\calF$ is complete, then clearly $\calF(a,w) \neq \es$ for all $a\in A$ and $w\in X^*$.


\section{Directing words and directable automata}\label{se:DirWords}

A word $w\in X^*$ is a \emph{directing word} of a DFA $\faA$ if there is a state $c\in A$ such that $\delta^*(a,w) = c$ for every $a\in A$. The set of directing words of $\calA$ is denoted by $DW(\calA)$, and $\calA$ is \emph{directable} if $DW(\calA) \neq \es$.

In \cite{ImSt99} three kinds of directing words of an NFA $\ndN$ were identified. A word $w\in X^*$ was said to be
\begin{itemize} \vspace{-3pt}
  \item[{\rm (D1)}] \emph{D1--directing} if $(\exists c\in A)(\forall a\in A)(\alpha^*(a,w) = \{ c\} )$, \vspace{-3pt}
  \item[{\rm (D2)}] \emph{D2--directing} if $(\forall a,b\in A)(\alpha^*(a,w) = \alpha^*(b,w)$, and \vspace{-3pt}
  \item[{\rm (D3)}] \emph{D3--directing} if $(\exists c\in A)(\forall a\in A)(c\in \alpha^*(a,w))$. \vspace{-3pt}
\end{itemize}
For each $i=1,2,3$, the set of  D$i$--directing words of $\calN$ is denoted by $D_i(\calN)$, and $\calN$ is called D$i$--\emph{directable} if $D_i(\calN)\neq \es$.

If we regard a DFA as a special NFA, all three conditions (D$i$) yield the usual directing words. To state this more precisely, we associate with each DFA $\faA$ an NFA $\An =(A,X,\alpha)$, where $\alpha : A \times X \ra \wp(A)$ is defined by $\alpha(a,x) := \{\delta(a,x)\}$ ($a\in A, x \in X$). It is clear that $\alpha^*(a,w) = \{\delta^*(a,w)\}$ for all $a\in A$ and $w \in X^*$, which implies the following facts.

\begin{remark}\label{re:DW and Di of a DFA}
For any DFA $\calA$, $DW(\calA) = D_1(\An) = D_2(\An) = D_3(\An)$.
\end{remark}

Clearly, the directability of fuzzy automata may also be defined in several ways. The definition suggested by \text{Karthikeyan} and \text{Rajasekar} \cite{KaRa15} may be stated as follows: $w\in X^*$ is a \emph{directing word} of an FFA $\fzF$ if there is a state $c\in A$ such that $f^*(a,w,c) > 0$ (i.e. $c \in \calF(a,w)$) for every $a\in A$, and   $\calF$ is said to be \emph{directable} if it has a directing word. Using these notions the authors attempt to present some ideas and results appearing in \cite{ImSt95} in a `fuzzified' form. The discussion of these and related matters can be simplified by the following reduction to nondeterministic automata.

Consider an FFA $\fzF$. A word $w = x_1\ldots x_k$ $(x_1,\ldots,x_k\in X)$ is directing for $\calF$ if and only if there is a state $c\in A$ such that for every $a\in A$ there is a chain of states $a = a_0,a_1,\ldots,a_k = c$ such that $f(a_{i-1},x_i,a_i) > 0$ for every $i \in \bfk$. The actual values of the degrees $f(a_{i-1},x_i,a_i)$ do not matter as long as they are positive. Hence, we could assign every positive transition the value 1 without changing the set of directing words. This means that when we are interested just in directing words, $\calF$ may be replaced by the \emph{associated NFA} $\Fnda$, where $\alpha$ is defined by \vspace{-3pt}
\[\alpha(a,x) := \{b\in A \mid f(a,x,b) > 0\} \qquad ( x\in X, a\in A).\vspace{-3pt}
\]

\begin{example}\label{ex:Fuzzy to NFA}{\rm Let $\calF = (\{a,b,c\},\{x,y\},f)$ be the fuzzy automaton, where the non-zero transitions are $f(a,x,b) = 0.3$, $f(b,x,c) = 0.4$, $f(c,x,b) = 0.2$, $f(c,x,c) = 0.6$, $f(b,y,b) = 0.5$ and $f(b,y,c) = 0.1$.
Then the transitions of $\Fnd$ are given by $\alpha(a,x) = \{b\}$, $\alpha(b,x) = \{c\}$, $\alpha(c,x) = \{b,c\}$, $\alpha(a,y) = \es$, $\alpha(b,y) = \{b,c\}$ and $\alpha(c,y) = \es$.}
\end{example}

It is easy to verify the following lemma by induction on $\lg(w)$.

\begin{lemma}\label{le:fzA to ndA} Let $\fzF$ be  any FFA and $\Fnda$ be the associated NFA. Then $\alpha^*(a,w) = \calF(a,w)$
for all $a\in A$ and $w\in X^*$.
\end{lemma}

Let us also note the following obvious fact.

\begin{corollary}\label{co:Completeness of automata} A fuzzy finite automaton $\calF$ is complete if and only if the associated NFA $\Fnd$ is complete.
\end{corollary}

\begin{remark}\label{re:NFA to FFA} A converse construction is possible, too, i.e., with any NFA $\ndN$, we may associate a related FFA $\Nfza$, where $f : A \times X \times A \ra [0,1]$ is defined by
\[
f(a,x,b) = \left\{\begin{array}{cl}
                         1,&\ \mbox{if } b \in \alpha(a,x)\\
                         0,&\  \mbox{if } b \notin \alpha(a,x). \vspace{-3pt}
                         \end{array}
                   \right.
\]
It is easy to see that $\Nfz$ is a crisp FFA, and that

{\rm (a)} $\Nfz$ is complete if and only if $\calN$ is complete,

{\rm (b)} $\Nfz(a,w) = \alpha^*(a,w)$ for all $a \in A$ and $w \in X^*$, and

{\rm (c)} $(\Nfz)^{\bfn\bfd} = \calN$.\\
However, $(\Fnd)^{\bff\bfz} = \calF$ holds for an FFA $\calF$ only in case $\calF$ is crisp.
\end{remark}

The above definitions of D1--, D2-- and D3--directing words can be applied in a natural way to fuzzy automata, too.
For an FFA $\fzF$, let us call a word $w\in X^*$
\begin{itemize} \vspace{-3pt}
  \item[{\rm (D1')}] \emph{D1--directing} if $(\exists c\in A)(\forall a\in A)(\calF(a,w) = \{c\} )$, \vspace{-3pt}
  \item[{\rm (D2')}] \emph{D2--directing} if $(\forall a,b\in A)(\calF(a,w) = \calF(b,w))$, and\vspace{-3pt}
  \item[{\rm (D3')}] \emph{D3--directing} if $(\exists c\in A)(\forall a\in A)(c \in \calF(a,w))$. \vspace{-3pt}
\end{itemize}
For each $i=1,2,3$, the set of  D$i$--directing words of $\calF$ is denoted by $D_i(\calF)$, and $\calF$ is said to be  D$i$--\emph{directable} if $D_i(\calF)\neq \es$.

The D3--directing words of a fuzzy automaton $\calF$ are exactly the directing words as defined  in \cite{KaRa15}.
A pairwise comparison of conditions (Di) and (Di') ($i=1,2,3$)  yields by Lemma \ref{le:fzA to ndA} the following facts.

\begin{proposition}\label{pr:Di(fzA) = Di(ndA)} For any FFA $\calF$ and each $i = 1,2,3$, $D_i(\calF) = D_i(\Fnd)$, and hence
$\calF$ is Di-directable if and only if $\Fnd$ is Di-directable.
\end{proposition}

Proposition \ref{pr:Di(fzA) = Di(ndA)} implies that all the known results  concerning D$i$--directing words of NFAs  (cf. \cite{Bur76,GaIvNG09,GoHeKoRy82,ImSt99,ItST04,Sha18}, for example) apply to fuzzy automata, too. For example,
the following relations between the sets $D_1(\calF)$, $D_2(\calF)$, and $D_3(\calF)$, for any fuzzy automaton $\calF$, hold by Remark 3.2 of \cite{ImSt99}.

\begin{remark}\label{re:D1F,D2F and D3F} $D_1(\calF) \se D_2(\calF)\cap D_3(\calF)$ for any fuzzy automaton $\calF$, and if $\calF$ is complete, then $D_1(\calF) \se D_2(\calF) \se D_3(\calF)$. Moreover, any one of the inclusions may be proper.
\end{remark}

As noted in \cite{ImSt95}, for example,  $X^*DW(\calA)X^* = DW(\calA)$ for any DFA $\calA$,  but the corresponding fact does not always hold for the D3--directing words of a fuzzy automaton $\calF$ as claimed in \cite{KaRa15}. In fact, neither $X^*D_3(\calF) = D_3(\calF)$ nor $D_3(\calF)X^* = D_3(\calF)$ holds generally. For example, the fuzzy automaton $\calF$ of Example \ref{ex:Fuzzy to NFA} has the D3--directing word $xx$ while neither $yxx$ nor $xxy$ is in $D_3(\calF)$. Remark 3.3 of \cite{ImSt99} yields the following facts.

\begin{remark}\label{re:X*DW(F)X* = DW(F)} $D_2(\calF)X^* = D_2(\calF)$ for any FFA $\calF$. If $\calF$ is complete, then $X^*D_1(\calF) = D_1(\calF)$, $X^*D_2(\calF)X^* = D_2(\calF)$ and $X^*D_3(\calF)X^* = D_3(\calF)$.
\end{remark}

It is rather obvious that  $D_1(\calF)$, $D_2(\calF)$ and $D_3(\calF)$ are regular languages for every FFA $\calF$. Although this also follows directly from Proposition 3.4 of \cite{ImSt99}, let us present a simple proof formulated for FFAs. 

\begin{proposition}\label{pr:DW(A) regular} For any  FFA $\calF$, the languages $D_1(\calF)$, $D_2(\calF)$ and $D_3(\calF)$ are regular.
\end{proposition}

\bp If $\fzF$ with $A = \{a_1,\ldots,a_n\}$, let $(B,X,\delta)$ be the DFA with $B=\{\{\calF(a_1,u),\dots ,\calF(a_n,u)\} \mid u\in X^*\}$ and $\delta$ defined by \vspace{-3pt}
$$
\delta(\{C_1,\dots,C_k\},x)= \{\calF(C_1,x),\dots,\calF(C_k,x)\}\vspace{-3pt}
$$
for all $\{C_1,\dots,C_k\} \in B$ and $x\in X$. 
For $b_0=\{\{ a_1\},\dots,\{a_n\}\}$, we get
$\delta(b_0,u)=\{\calF(a_1,u),\dots,\calF(a_n,u)\}$ for every $u\in X^*$. It is easy to see that, for each $i =1,2,3$, $L(\bfB_i)= D_i(\calF)$
for $\bfB_i =(B,X,\delta,b_0,F_i)$, when we choose $F_1 = \{\{\{c\}\} \mid c\in A\}\cap B$, $F_2 = \{\{C\} \mid C \se A\}\cap B$, and
$F_3 =\{\{C_1,\dots,C_k\}\in B \mid C_1\cap\dots\cap C_k\neq \es \}$.
\ep

The construction used in the above proof is clearly effective if $\calF$ is effectively given. This means that we may decide whether $\calF$ is D1--, D2-- or D3--directable. However, the decision method suggested by this observation is feasible for  very small FFAs only. We shall now present a more effective way to decide the D3--directability of a complete FFA.\smallskip

Two states $a$ and $b$ of a DFA $\faA$ are \emph{merged} by a word $w\in X^*$ if $\delta^*(a,w) = \delta^*(b,w)$. The algorithm for testing the directability of a DFA $\calA$ presented in \cite{ImSt95} is based on the obvious fact that  $\calA$ is directable if and only if every pair of states of $\calA$ can be merged.

By the definition suggested in \cite{KaRa15}, two states $a$ and $b$ of an FFA $\fzF$ are \emph{merged} by a word $w\in X^*$ if $f^*(a,w,c) >0$ and $f^*(b,w,c) >0$ for some $c\in A$. Theorem 4.1 of \cite{KaRa15} claims that $\calF$ is directable if and only if every pair of states can be merged. The condition is trivially necessary, but the converse may fail if the automaton is incomplete, and then the algorithm of \cite{KaRa15}, obtained from that of \cite{ImSt95} by replacing the Boolean matrix by a fuzzy matrix, may incorrectly proclaim the automaton to be directable.

\begin{example}\label{ex:Dir and merging in fzA}{\rm Let $\calF = (\{a,b,c\},\{x,y,z\},f)$ be the FFA in which $f(a,x,a)$, $f(b,x,a)$, $f(b,y,b)$, $f(c,y,b)$, $f(a,z,c)$ and $f(c,z,c)$ are the transitions with non-zero degrees.
Clearly, $a$ and $b$ are merged by $x$, $b$ and $c$ by $y$, and $a$ and $c$ by $z$, but  $D_3(\calF) = \es$ since $f^*(a,yw,d) = f^*(b,zw,d) = f^*(c,xw,d) = 0$ for all $w\in X^*$ and $d\in A$.
}
\end{example}

However, the algorithm of \cite{ImSt95} can be used, almost without any change, for testing a complete FFA for D3--directability.

In \cite{ImSt99}  a word $w\in X^*$ was defined to \emph{D3--merge} two states $a,b\in A$ of an NFA $\ndN$ if $\alpha^*(a,w)\cap \alpha^*(b,w) \neq \es$. Let us say that $w$ \emph{D3--merges} two states $a,b\in A$ of an FFA $\calF$ if $\calF(a,w)\cap\calF(b,w) \neq \es$. A word merges, in the sense of \cite{KaRa15}, two states if and only if it D3--merges them.

\begin{lemma}\label{le:Dir and merge in fzA} A word D3--merges two states of an FFA $\calF$ if and only if it D3--merges them in the associated NFA $\Fnd$. Hence a complete FFA is D3--directable if and only if every pair of its states has a D3--merging word.
\end{lemma}

\bp The first assertion is valid by Lemma \ref{le:fzA to ndA}.  The rest of the lemma follows then from Lemma 5.4 of \cite{ImSt99} which expresses the corresponding fact for NFAs. \ep

For any $k\geq 0$, two states $a,b\in A$ of an FFA  $\fzF$ are said to be \emph{D3 $k$-mergeable} if they are D3--merged by a word of length $\leq k$. We  denote the D3 $k$-mergeability relation of $\calF$ by  $\mu(k)$, and let $\mu := \bigcup_{k\geq 0}\mu(k)$.

It is easy to verify the following proposition that justifies the algorithm to be presented. It corresponds exactly to  Proposition 4.1 of \cite{ImSt95}.

\begin{proposition}\label{pr:Mu-relations} Let $\fzF$ be a complete FFA.
\begin{itemize} \vspace{-3pt}
  \item[{\rm (a)}] $\calF$ is D3--directable if and only if $\mu = A\times A$. \vspace{-3pt}
  \item[{\rm (b)}] The relations $\mu(k)$ and $\mu$ are reflexive and symmetric. \vspace{-3pt}
  \item[{\rm (c)}] $\mu(0) = \Delta_A$, and $\mu(k) = \mu(k-1) \cup\\  \{(a,b) \mid (\exists x\in X)[(\calF(a,x) \times \calF(b,x))\cap \mu(k-1) \neq \es]\}$ for $k\geq 1$. \vspace{-3pt}
  \item[{\rm (d)}] If $\mu(k) = \mu(k-1)$ for some $k\geq 1$, then $\mu(k) = \mu$. \vspace{-3pt}
  \item[{\rm (e)}] $\Delta_A = \mu(0) \subset \mu(1) \subset \ldots \subset \mu(k) = \mu$ for some $k \leq {n \choose 2}$, where $n = |A|$. \vspace{-3pt}
\end{itemize}
\end{proposition}

Let $\fzF$ be an $n$-state complete FFA with  $A = \{1,\ldots,n\}$. The algorithm implicitly computes the relations $\mu(0), \mu(1), \ldots$ until $\mu(k-1) = \mu(k)$ for some $k\geq 1$. It employs a Boolean $n\times n$-matrix $M$ and a list $NewPair$ of pairs of states. The meaning of $M[i,j] = 1$ is that states $i$ and $j$ are known to be D3--mergeable. It suffices to consider the entries $M[i,j]$ with $1\leq i < j \leq n$. A pair $(i,j)$ appears on the list $NewPair$ if it is known to be D3--mergeable but this fact has not yet been utilized for finding new D3--mergeable pairs. We also use the \emph{inverted transition table} $\bfI = (\bfI[a,x])_{a\in A,x\in X}$, where $\bfI[a,x] := \{i\in A \mid a\in \calF(i,x)\}$ for all $a\in A$ and $x\in X$.
The steps of the algorithm are as follows:

1.  Set $M[i,j] := 0$ for $1\leq i<j \leq n$, $NewPair := \ve$ (empty list), and compute $\bfI$.

2. Find all pairs $(a,x)\in A\times X$ for which $|\bfI[a,x]| > 1$. For each such pair $(a,x)$ consider every pair $i,j \in \bfI[a,x]$ with $i<j$. If $M[i,j] = 0$, let $M[i,j] := 1$ and append $(i,j)$ to $NewPair$.

3. Until $NewPair = \ve$ do the following. Delete the first pair, say $(a,b)$, from $NewPair$. For each $x\in X$, find all pairs $(i,j)$ with $i<j$ such that $i \in \bfI[a,x]$ and $j\in \bfI[b,x])$, or $j \in \bfI[a,x]$ and $i\in \bfI[b,x])$. If $M[i,j] = 0$, let $M[i,j] := 1$ and append $(i,j)$ to $NewPair$.

4. $\calF$ is D3--directable if and only if $M[i,j] = 1$ whenever $1\leq i < j \leq n$.

Quite the same way as in \cite{ImSt95}, one can show that the time bound for the algorithm is $\calO(m\cdot n^2)$, where $m = |X|$ and $n= |A|$.

It appears unlikely that anything like the above algorithm exists for D1-- or D2--directability as there does not seem to be any useful notions of D1-- or D2--merging words. Furthermore, while the shortest D3--directing word of a  complete D3--directable  $n$-state NFA is at most of length of order $\mathcal{O}(n^3)$ (cf. Proposition 5.3 of \cite{ImSt99}), no polynomial upper bounds exist for shortest D1-- and D2--directing words (cf. \cite{Bur76,GaIvNG09,GoHeKoRy82,ImSt99}).


\section{Directability with degrees}

We shall now introduce three new types of directing words for FFAs that are analogous to the D1--, D2-- and D3--directing words but depend more on the transition degrees $f(a,x,b)$ of the automaton.

For a given FFA $\fzF$, we say that a word $w\in X^*$ is
\begin{itemize} \vspace{-3pt}
  \item[{\rm (DD1)}] \emph{DD1--directing} if $(\exists c\in A)(\exists r \in (0,1])(\forall a\in A)(\calF_{a,w} = \{c/r\} )$, \vspace{-3pt}
  \item[{\rm (DD2)}] \emph{DD2--directing} if $(\forall a,b\in A)(\calF_{a,w} = \calF_{b,w})$, and\vspace{-3pt}
  \item[{\rm (DD3)}] \emph{DD3--directing} if $(\exists c\in A)(\forall a\in A)(\calF_{a,w}(c) = \max\{\calF_{a,w}(b) \mid b \in A\} >0)$. \vspace{-3pt}
\end{itemize}
For each $i = 1,2,3$, the set of DD$i$--directing words of $\calF$ is denoted by $D\!D_i(\calF)$, and $\calF$ is said to be \emph{DD$i$--directable} if
$D\!D_i(\calF) \neq \es$.

If condition (DD1) holds, then for every state $a\in A$, $c$ is the only state reachable from $a$ by the input word $w$, and its reachability degree $\calF_{a,w}(c)$ has the same value $r>0$ for every $a$.
That $w$ satisfies (DD2) means that the fuzzy set $\calF_{a,w}$ of states reached from a given state $a$ by input $w$ is the same for every $a$, i.e., $f^*(a,w,c) = f^*(b,w,c)$ for all $a,b,c\in A$. Finally, condition (DD3) means that there is a state $c$ which has the maximal reachability degree $\calF_{a,w}(c) >0$ by the input $w$ from any state $a\in A$.

Let us see how these notions are related to each other and the previously considered types of directing words of fuzzy automata.

\begin{proposition}\label{pr:D$i$ and DD$i$}
Let $\fzF$ be any FFA.
\begin{itemize} \vspace{-3pt}
  \item[{\rm (a)}] $D\!D_1(\calF) \se D_1(\calF)$, \vspace{-3pt}
  \item[{\rm (b)}]  $D\!D_1(\calF) \se D\!D_2(\calF) \se D_2(\calF)$, \vspace{-3pt}
  \item[{\rm (c)}] $D\!D_1(\calF) \se D\!D_3(\calF) \se D_3(\calF)$, and \vspace{-3pt}
  \item[{\rm (d)}] $D\!D_1(\calF) \se D\!D_2(\calF) \cap D\!D_3(\calF)$. \vspace{-3pt}
\end{itemize}
Any one of the inclusions may be proper.
\end{proposition}

\bp
The inclusions follow rather easily from the definitions.  Let us prove statements (b) and (d).

If $w \in D\!D_1(\calF)$, and (DD1) holds for $c\in A$ and $r \in (0,1]$, then $\calF_{a,w} = \{c/r\} = \calF_{b,w}$ for all $a,b\in A$, and thus $w \in D\!D_2(\calF)$.

If $w \in D\!D_2(\calF)$, then $\calF(a,w) = \supp(\calF_{a,w}) = \supp(\calF_{w,b}) = \calF(b,w)$ for all $a,b \in A$, and hence $w \in D_2(\calF)$.

To show that the inclusion $D\!D_1(\calF) \se D\!D_2(\calF)$ may be proper, we consider the FFA $\fzF$, where $A = \{a,b\}$, $X = \{x\}$, and $f$ is defined by $f(a,x,a) = f(b,x,a) = f(a,x,b) = f(b,x,b) = 1$. The word $x$ does not belong to $D\!D_1(\calF)$ because $\calF_{a,x} = \{a/1,b/1\}$. On the other hand, $x\in D\!D_2(\calF)$ because $\calF_{a,x} = \{a/1,b/1\} = \calF_{b,x}$.

The inclusion $D\!D_2(\calF) \se D_2(\calF)$ is proper for  $\calF = (\{a,b\},\{x\},f)$, if $f$ is defined by  $f(a,x,b) = 0.1$ and $f(b,x,b) = 0.2$. Now $x \notin D\!D_2(\calF)$ because $\calF_{a,x}(b) = 0.1$ while $\calF_{b,x}(b) = 0.2$. On the other hand, $x \in D_2(\calF)$ since $\calF(a,x) =\{b\} = \calF(b,x)$.

The inclusion relation in (d) follows from (b) and (c). That it may be proper, is seen by considering the first FFA defined above in this proof; the word $x$ belongs to $D\!D_2(\calF) \cap D\!D_3(\calF)$, but not to $D\!D_1(\calF)$
\ep
\smallskip

Let us say that an FFA $\fzF$ is \emph{crisp} if $f(a,x,b) \in \{0,1\}$ for all $a,b \in A$ and $x \in X$. It is easy to see that if $\calF$ is crisp, then for all $a,b \in A$ and $w \in X^*$,

(1) $f^*(a,w,b) \in \{0,1\}$, and

(2) $\calF_{a,w} = \{a_1/1,\ldots,a_k/1\}$, where $\{a_1, \ldots,a_k\} = \calF(a,w)$.

From (2) we get the following facts.

\begin{remark}\label{re:Crisp FFAs and NFAs}
If $\fzF$ is a crisp FFA, then $D\!D_1(\calF) = D_1(\Fnd) = D_1(\calF)$,  $D\!D_2(\calF) = D_2(\Fnd) = D_2(\calF)$ and $D\!D_3(\calF) = D_3(\Fnd) = D_3(\calF)$.
\end{remark}

\bp Let us verify the first pair of equalities. For any $w \in X^*$, \vspace{-3pt}
\begin{align*}
w \in D\!D_1(\calF) &\LRa (\exists c \in A)(\exists r \in (0,1])(\forall a \in A) \calF_{a,w} = \{c/r\}\vspace{-3pt}\\
&\LRa (\exists c \in A)(\forall a \in A) \calF_{a,w} = \{c/1\}\vspace{-3pt}\\
&\LRa (\exists c \in A)(\forall a \in A) \calF(a,w) = \{c\}\vspace{-3pt}\\
&\LRa w \in D_1(\Fnd). \vspace{-3pt}
\end{align*}
Hence, $D\!D_1(\calF) = D_1(\Fnd)$, and $D_1(\Fnd) = D_1(\calF)$ by Proposition \ref{pr:Di(fzA) = Di(ndA)}.
The other two pairs of equalities have similar straightforward proofs.
\ep
\smallskip

When dealing with fuzzy finite automata it may be convenient to use their matrix representations.
For any $m,n \geq 1$, a \emph{fuzzy $m\times n$-matrix} is an $m\times n$-array $(r_{ij})$ in which $r_{ij} \in [0,1]$ for all $i \in \bfm$ and $j \in \bfn$. The \emph{product} of a fuzzy $m \times k$-matrix $R = (r_{ij})$ and a fuzzy $k \times n$-matrix $S = (s_{ij})$ is the fuzzy $m \times n$-matrix $RS = (t_{ij})$, where \vspace{-3pt}
\[
t_{ij} = (r_{i1}\land s_{1j}) \vee \ldots \vee (r_{ik}\land s_{kj}) \qquad (i\in \bfm, j\in \bfn).\vspace{-3pt}
\]

Let $\fzF$ be an $n$-state FFA with $A = \{a_1,\ldots,a_n\}$. With any word $w \in X^*$, we associate its fuzzy $n \times n$ \emph{transition matrix} \vspace{-3pt}
\[
M^\calF_w = M_w = (r_{ij}), \text{ where } r_{ij} = f^*(a_i,w,a_j) \qquad (i,j \in \bfn).
\]
\vspace{-3pt}
For any words $u,v \in X^*$, $M_{uv} = M_uM_v$ because
\vspace{-3pt}
\[
f(a_i,uv,a_j) = \max\{f^*(a_i,u,a_k)\land f^*(a_k,v,a_j) \mid k \in \bfn\} \qquad (i,j \in \bfn). \vspace{-3pt}
\]
For any $w\in X^*$ and $i \in \bfn$, the $i^{th}$ row $(r_{i1}\, r_{i2}\, \ldots \, r_{in})$ of $M_w$ represents the fuzzy set $\calF_{a_i, w}$, i.e.,  \vspace{-3pt}
\[
r_{i1} = \calF_{a_i, w}(a_1), \,r_{i2} = \calF_{a_i, w}(a_2), \ldots, r_{in} = \calF_{a_i, w}(a_n) \qquad  (i \in \bfn) \vspace{-3pt}
\]
By this observation, one may deduce from the matrix $M_w$ whether the word $w$ is DD1--, DD2-- or DD3--directing, and this will be used in the proof of the following proposition.

\begin{proposition}\label{pr:DFi(F) recognizable}
For any FFA $\fzF$, the languages $D\!D_1(\calF)$, $D\!D_2(\calF)$ and $D\!D_3(\calF)$ are recognizable.
\end{proposition}

\bp Let $A = \{a_1,\ldots,a_n\}$. The set $B := \{M_w \mid w \in X^*\}$ is finite because all the matrix entries belong to the finite set $\{f(a_i,x,a_j) \mid i,j \in \bfn, x\in X\} \cup \{1\}$. To see this, it suffices to notice that no new elements are introduced when matrices are multiplied as the only operations used are the join $\max(r,s)$ and meet $\min(r,s)$ in the chain $([0,1],\leq)$.

Let $(B,X,\delta)$ be the DFA for which $\delta : B \times X \ra B$ is defined by $\delta(M_w,x) = M_{wx}$ ($w\in X^*, x\in X$). It is easy to show that $\delta^*(M_\ve,w) = M_w$ for every $w \in X^*$.
From this observation, it follows that for each $i = 1,2,3$, we get a DFR $\bfB_i = (B,X,\delta,M_\ve,F_i)$ recognizing $D\!D_i(\calF)$ by the following choices of the sets $F_i$ of final states. \vspace{-3pt}
\begin{enumerate}
  \item $M_w\in F_1$ if and only if for some $j \in \bfn$ and $r \in (0,1]$, $r_{1j} = \ldots = r_{nj} = r$ and $r_{ik} = 0$ for all $i,k\in \bfn$ with $k \neq j$.\vspace{-3pt}
  \item $M_w \in F_2$ if and only if all the rows in $M_w$ are identical.\vspace{-3pt}
  \item $M_w\in F_3$ if and only if for some $j \in \bfn$, in every row, the $j^{th}$ element is maximal and positive. \ep
\end{enumerate}

In Remark \ref{re:X*DW(F)X* = DW(F)} we noted some equalities satisfied by the $D_i(\calF)$-languages, but not all of them hold for the $D\!D_i(\calF)$-languages. For example, let $\calF = (\{a,b\},\{x,y\},f)$ be the complete FFA in which the nonzero transitions are $f(a,x,b) = f(b,x,b) = f(b,y,b) = 1$ and $f(a,y,a) = 0.5$. Then $x \in D\!D_1(\calF)$ because $\calF_{a,x} = \calF_{b,x} = \{b/1\}$, but $yx \notin D\!D_1(\calF)$ since $\calF_{a,yx} = \{b/0.5\}$ while $\calF_{b,yx} = \{b/1\}$. Hence, $X^*D\!D_1(\calF) \neq D\!D_1(\calF)$. Similarly, $X^*D\!D_2(\calF) \neq D\!D_2(\calF)$ because $x$ is a D2--directing word of $\calF$, but $yx$ is not.
However, the equality $X^*D\!D_1(\calF) = D\!D_1(\calF)$, for example, holds if we replace the completeness of an FFA by the stronger property of ``normality''.
 
Let us say that an FFA $\fzF$ is \emph{normal} if for every pair $a\in A, \, x\in X$, there is a state $b \in A$ for which $f(a,x,b) = 1$, i.e., $\calF_{a,x}$ is a normal fuzzy set. 

\begin{lemma}\label{le:normal for words}
If $\fzF$ is a normal FFA, then for all $a \in A$ and $w \in X^*$, there is a state $b\in A$ such that $f^*(a,w,b) = 1$. Hence, if $\calF$ is normal and $w \in D\!D_3(\calF)$, then there is a $c \in A$ such that $f^*(a,w,c) = 1$ for every $a\in A$.
\end{lemma}

\begin{proposition}\label{pr:X*DDi-laws}
For any FFA $\fzF$, \vspace{-3pt}
\begin{itemize}
  \item[{\rm (a)}] $D\!D_2(\calF)X^* = D\!D_2(\calF)$.  \vspace{-3pt}
\end{itemize}
If $\calF$ is normal, then we also have  \vspace{-3pt}
\begin{itemize}
  \item[{\rm (b)}] $X^*D\!D_1(\calF) = D\!D_1(\calF)$,
  \item[{\rm (c)}] $X^*D\!D_2(\calF)X^* = D\!D_2(\calF)$, and
  \item[{\rm (d)}] $X^*D\!D_3(\calF)X^* = D\!D_3(\calF)$.
\end{itemize}
\end{proposition}

\bp To prove (a), consider any words $w \in D\!D_2(\calF)$ and $u \in X^*$. For all $a,b,c \in A$, \vspace{-3pt}
\begin{align*}
     \calF_{a,wu}(c) &=  \max\{f^*(a,w,d) \land f^*(d,u,c) \mid d \in A\}\vspace{-3pt}\\
        &=  \max\{f^*(b,w,d) \land f^*(d,u,c) \mid d \in A\}
        = \calF_{b,wu}(c),\vspace{-3pt}
\end{align*}
which shows that $D\!D_2(\calF)X^* \se D\!D_2(\calF)$. The converse inclusion is obvious.

Assume now that $\calF$ is normal. To prove (b), consider any  $w \in D\!D_1(\calF)$, and let $c \in A$ and $r \in (0,1]$ be such that $\calF_{a,w} = \{c/r\}$ for every $a \in A$. Take any $u \in X^*$ and $a \in A$. By Lemma \ref{le:normal for words}, there is a $b \in A$ such that $f^*(a,u,b) = 1$, and therefore \vspace{-3pt}
\begin{align*}
f^*(a,uw,c) &= \max\{f^*(a,u,d) \land f^*(d,w,c) \mid d \in A\}\vspace{-3pt}\\
&= \max\{f^*(a,u,d) \mid d \in A\} \land r\vspace{-3pt}\\ 
&= 1 \land r = r,\vspace{-3pt}
\end{align*}
where we used the fact that $f^*(d,w,c) = r$ for every $d \in A$. On the other hand, if $c' \in A$, $c' \neq c$, then for every $a \in A$\vspace{-3pt}
\begin{align*}
f^*(a,uw,c') &= \max\{f^*(a,u,d) \land f^*(d,w,c') \mid d \in A\}\vspace{-3pt}\\
&= \max\{f^*(a,u,d) \land 0 \mid d\in A\}\vspace{-3pt} = 0.\vspace{-3pt}
\end{align*}This shows that $uw \in D\!D_1(\calF)$, and hence $X^*D\!D_1(\calF) \se D\!D_1(\calF)$. 

As we already have $D\!D_2(\calF)X^* = D\!D_2(\calF)$, we get (c) by showing that $X^*D\!D_2(\calF) \se D\!D_2(\calF)$. Let $u \in X^*$ and $w \in D\!D_2(\calF)$, and consider any $a,b,c\in A$. Let $a',b'\in A$ be such that $f^*(a,u,a') = f^*(b,u,b') = 1$. Then \vspace{-3pt}
\begin{align*}
f^*(a,uw,c) &= \max\{f^*(a,u,d) \land f^*(d,w,c) \mid d \in A\}\vspace{-3pt}\\
&= f^*(a,u,a') \land f^*(a',w,c)\vspace{-3pt}\\
&= 1 \land f^*(a',w,c) = 1 \land f^*(b',w,c)\vspace{-3pt}\\
&= \max\{f^*(b,u,d) \land f^*(d,w,c) \mid d \in A\}\vspace{-3pt}\\ 
&= f^*(b,uw,c),\vspace{-3pt}
\end{align*}
where we also used the fact that $f^*(a',w,c) = f^*(d,w,c) = f^*(b',w,c)$ for every $d \in A$. Thus we have shown that $uw \in D\!D_2(\calF)$.

Finally, to prove (d), consider any $u \in X^*$ and $w \in D\!D_3(\calF)$. Let $c\in A$ be a state such that, for every $a\in A$, $f^*(a,w,c) >0$ and $f^*(a,w,b) \leq f^*(a,w,c)$ for every $b \in A$. If $c' \in A$ is a state such that $f^*(c,u,c') = 1$, then for any $a\in A$, \vspace{-3pt}
\begin{align*}
f^*(a,wu,c') &= \max\{f^*(a,w,d) \land f^*(d,u,c') \mid d \in A\}\vspace{-3pt}\\ 
&\geq f^*(a,w,c) \land f^*(c,u,c')\vspace{-3pt}\\ 
&= f^*(a,w,c) > 0.
\end{align*}
Moreover, for any $b \in A$,\vspace{-3pt}
\begin{align*}
f^*(a,wu,b) &= \max\{f^*(a,w,d) \land f^*(d,u,b) \mid d \in A\}  \vspace{-3pt}\\
&\leq f^*(a,w,c) \land \max\{f^*(d,u,b) \mid d \in A\}\vspace{-3pt}\\
&\leq f^*(a,w,c) = f^*(a,w,c) \land f^*(c,u,c')\vspace{-3pt}\\
&\leq f^*(a,wu,c'),
\end{align*}
where we used the fact that $f(a,w,d) \leq f(a,w,c)$ for every $d \in A$, 
and thus $wu \in D\!D_3(\calF)$. The converse inclusion $D\!D_3(\calF) \se D\!D_3(\calF)X^*$ holds trivially.
To complete the proof of (d), we show that $X^*D\!D_3(\calF) \se D\!D_3(\calF)$. Let $w\in D\!D_3(\calF)$, $c \in A$ and $u \in X^*$ be as above. Consider any $a \in A$ and let $a' \in A$ be a state for which $f^*(a,u,a') = 1$. Then
\vspace{-3pt}
\begin{align*}
f^*(a,uw,c) &= \max\{f^*(a,u,d) \land f^*(d,w,c) \mid d \in A\}  \vspace{-3pt}\\
&\geq f^*(a,u,a') \land f^*(a',w,c) = f^*(a',w,c) > 0\vspace{-3pt}
\end{align*}
Moreover, for any $b \in A$,
\vspace{-3pt}
\begin{align*}
f^*(a,uw,b) &= \max\{f^*(a,u,d) \land f^*(d,w,b) \mid d \in A\}  \vspace{-3pt}\\
&\leq \max\{f^*(a,u,d) \land f^*(d,w,c) \mid d \in A\} = f^*(a,uw,c), \vspace{-3pt}
\end{align*}
and hence $uw \in D\!D_3(\calF)$.
\ep

It is easy to see that if  $L = RX^* (\se X^*)$, where $R$ is a nonempty regular language, then any minimal DFR $\frA$ of $L$ has exactly one final state, and this state $d$ is a \emph{trap state}, i.e., $\delta(d,x) = d$ for every $x \in X$. Hence, Proposition \ref{pr:X*DDi-laws} yields the following facts.

\begin{corollary}\label{co:Minimal DFR with traps}
If an FFA $\fzF$ is $D\!D2$-directable, then any minimal DFR recognizing $D\!D_2(\calF)$ has exactly one final state, and this is a trap state. If $\calF$ is normal and D3--directable, then any minimal DFR recognizing $D\!D_3(\calF)$ has exactly one final state, and this is a trap state.
\end{corollary}


\section{Families of sets of $DDi$-directing words}\label{se:Families}

Let us now consider the families of languages $D\!D_i(\calF)$ formed by the DD$i$--directing words of DD$i$--directable FFAs.
In \cite{ImIt99} Imreh and Ito studied the families of the sets $D_i(\calN)$ of D$i$--directing words of $Di$-directable NFAs. Their results, also presented in \cite{Ito04}, apply directly to fuzzy automata, too. 

Let $\flL$ and $\flK$ be families of languages. $\calK$ is a \emph{subfamily} of $\calL$, $\calK \se \calL$ in symbols, if $\calK(X) \se \calL(X)$ for every alphabet $X$. If $\calK \se \calL$, but $\calK \neq \calL$, then $\calK$ is a \emph{proper subfamily} of $\calL$ and we write $\calK \subset \calL$. The \emph{intersection} of $\calK$ and $\calL$ is the family $\calK \cap \calL := \{\calK(X) \cap \calL(X)\}_X$. The families $\calK$ and $\calL$ are \emph{incomparable} if neither $\calK \se \calL$ nor $\calL \se \calK$ holds. This is expressed by $\calK \parallel \calL$.

For each $i \in \mathbf{3}$, let $\LDD_i = \{\LDD_i(X)\}_X$  be the family of the sets $D\!D_i(\calF)$ of DD$i$--directing words of DD$i$--directable FFAs. Similarly, let  $\nLDD_i = \{\nLDD_i(X)\}_X$ be the family of the sets $D\!D_i(\calF)$, where $\calF$ is a DD$i$--directable normal FFA.  Furthermore, let $\DW = \{\DW(X)\}_X$ be the family of the sets $DW(\calA)$ of directing words of directable DFAs. Note that any language belonging to $\LDD_1(X)$, $\LDD_2(X)$, $\LDD_3(X)$ or $\DW(X)$ for some $X$ is, by definition, nonempty.

The following facts are easily verified.

\begin{lemma}\label{le:DFA to FFA}
For any DFA $\faA$, the FFA $\Afz = (A,X,f)$ is normal, and for all $a,b \in A$ and $w \in X^*$,
\[
f^*(a,w,b) =  \left\{\begin{array}{cl}
                         1,&\ \mbox{if } \delta^*(a,w) = b\,;\\
                         0,&\  \mbox{otherwise}. \vspace{-3pt}
                         \end{array}
                   \right.
\]
Furthermore, $D\!D_1(\Afz) = D\!D_2(\Afz) = D\!D_3(\Afz) = DW(\calA)$.
\end{lemma}

\begin{corollary}\label{co:DW se nDD1 nDD2 nDD3}
$\DW \se \, n\LDD_1 \cap \nLDD_2 \cap n\LDD_3$.
\end{corollary}

It is clear that $X^*DW(\calA)X^* = DW(\calA)$ for any DFA $\calA$, and in \cite{ImIt99} it was shown that this is a characteristic property of the languages $DW(\calA)$:

\begin{proposition}\label{pr:Char of DW-sets}\,\rm{\cite{ImIt99}}\,
A language $L \se X^*$ belongs to $\DW(X)$  if and only if $L \in Rec(X)$, $L \neq \es$ and $X^*LX^* = L$.
\end{proposition}

\begin{proposition}\label{pr:DW sub DD1 cap DD2 cap DD3}
$\DW \subset \, \LDD_1 \cap \LDD_2 \cap \LDD_3$.
\end{proposition}

\bp The inclusion holds by Corollary \ref{co:DW se nDD1 nDD2 nDD3} (naturally $n\LDD_i \se \LDD_i$ for each $i$).
Let $\fzF$ be the FFA in which $A = \{a,b\}$, $X = \{x,y\}$. and $f$ is given by
$f(a,x,b) = f(b,x,b) = f(b,y,b) = 1$. Clearly, $\calF_{a,xu} = \calF_{b,xu} = \{b/1\}$ and $\calF_{a,yu} = \es$ for every $u \in X^*$. Thus $\LDD_1(\calF) = \LDD_2(\calF) = \LDD_3(\calF) = xX^*$. Since $xX^* \notin \DW(X)$, the claimed inclusion is proper.
\ep

\begin{proposition}\label{pr:DW, nDDi and DDi}
\begin{itemize}
  \item[{\rm (a)}] $\DW \subset n\LDD_1 \subset \LDD_1$.
  \item[{\rm (b)}] $\DW = n\LDD_2 \subset \LDD_2$.
  \item[{\rm (c)}] $\DW = n\LDD_3 \subset \LDD_3$.
\end{itemize}
\end{proposition}

\bp The relations $\DW \se n\LDD_i$ ($i = 1,2,3$) hold by Corollary \ref{co:DW se nDD1 nDD2 nDD3}. Let us prove separately the remaining parts of the three statements. 
\smallskip

To prove $\DW \subset n\LDD_1$, let us consider the normal FFA $\fzF$,  where $A = \{a,b\}$, $X = \{x,y\}$, and $f$ is given by
$f(a,x,b) = f(a,y,a) = f(b,x,b) = f(b,y,a) = f(b,y,b) = 1$. Clearly, $DD_1(\calF)$ is the language $X^*x$ which does not belong to $\DW(X)$.

That the inclusion $n\LDD_1 \subset \LDD_1$ is proper, is seen by considering the FFA $\fzF$,  where $A = \{a,b\}$, $X = \{x,y\}$, and $f$ is given by $f(a,x,b) = f(b,x,b)  = f(b,y,b) = 1$ and $f(a,y,a) = 0.5$. Obviously, $DD_1(\calF) = xX^*$, but $xX^* \notin n\LDD_1(X)$ by Proposition \ref{pr:X*DDi-laws}\,(b).
\smallskip

(b) To prove that $\DW = n\LDD_2$, it suffices to show that $n\LDD_2 \se \DW$. If $L \in n\LDD_2(X)$, then $L \neq \es$ and $L \in Rec(X)$. Moreover, $X^*LX^* = L$ by Proposition \ref{pr:X*DDi-laws}\,(c), and thus $L \in \DW(X)$

Let $\fzF$ be the FFA, where $A = \{a,b\}$, $X = \{x,y\}$ and $f$ is given by $f(a,x,b) = f(b,x,b) = f(b,y,b) = 1$. By Proposition \ref{pr:X*DDi-laws}\,(c), $DD_2(\calF) = xX^*$ is not in $n\LDD_2(X)$. Hence inclusion $n\LDD_2 \se \LDD_2$ is proper.

(c) That $\DW = n\LDD_3$ can be shown similarly as $\DW = n\LDD_2$ above. To prove  $n\LDD_3 \subset \LDD_3$, we consider again the FFA $\calF$ defined in part (b) above:  $DD_3(\calF^*) = xX^*$ is not in  $n\LDD_3(X)$.
\ep

\begin{proposition}\label{pr:nDD1 and DD2 equals nDD1 and DD3 equals DW}\vspace{-5pt}
\[
\DW =  n\LDD_1 \cap n\LDD_2 \cap n\LDD_3 = n\LDD_1 \cap \LDD_2 = n\LDD_1 \cap \LDD_3.
\]
\end{proposition}

\bp
The first equality follows from Corollary \ref{co:DW se nDD1 nDD2 nDD3} and Proposition \ref{pr:DW, nDDi and DDi}.

Also the inclusions $\DW \se n\LDD_1 \cap \LDD_2$ and $\DW \se n\LDD_1 \cap \LDD_3$ hold by Corollary \ref{co:DW se nDD1 nDD2 nDD3} and Proposition \ref{pr:DW, nDDi and DDi}.
On the other hand, if $L \in n\LDD_1 \cap \LDD_2$, then $L \neq \es$, $L \in Rec(X)$, and $X^*LX^* = L$ by Proposition \ref{pr:X*DDi-laws}\,(a,b). Thus $L \in \DW(X)$ by Proposition \ref{pr:Char of DW-sets}.
Hence,  $\DW = n\LDD_1 \cap \LDD_2$. The equality $\DW = n\LDD_1 \cap \LDD_3$ has a similar proof.
\ep

\begin{proposition}\label{pr:L_DD1 and L_DD3 incomparable}
$\LDD_1 \parallel \LDD_3$.
\end{proposition}

\bp There are two noninclusions to be shown.

If $\LDD_1  \se \LDD_3$, then naturally also $n\LDD_1 \se \LDD_3$. By Proposition \ref{pr:nDD1 and DD2 equals nDD1 and DD3 equals DW}, this would imply that $n\LDD_1 = n\LDD_1 \cap \LDD_3 = \DW$, which contradicts Proposition \ref{pr:DW, nDDi and DDi}\,(a). Thus $\LDD_1 \nsubseteq \LDD_3$. \smallskip

To prove $\LDD_3 \nsubseteq \LDD_1$, let $\fzF$ be the FFA, where $A = \{p,q\}$, $X = \{x,y\}$ and $f$ is defined by
\[
f(p,x,q) = f(q,x,p) = f(q,x,q) = f(p,y,p) =1.
\]
Then $x,xxy \in D\!D_3(\calF)$ as $\calF_{p,x} = \{q/1\}$, $\calF_{q,x} = \{p/1,q/1\}$, and $\calF_{p,xxy} = \{p/1\} = \calF_{q,xxy}$. Assume that there exists an FFA $\fzG$ for which $D\!D_1(\calG) = D\!D_3(\calF)$. Then there would exist $c,d \in B$ and $r,s \in (0,1]$ such that for every $a \in B$, $\calG_{a,x} = \{c/r\}$ and $\calG_{a,xxy} = \{d/s\}$. Thus, $\calG(a,x) = \{c\}$ for every $a\in B$, and $\calG(c,x) = \{c\}$ in particular. Since $\calG(a,xxy) = \{d\}$ for every $a \in B$, this must means also that $\calG(c,y) = \{d\}$. Moreover, for every $a \in B$, $g(a,x,c) =c$ and $g(a,x,b) = 0$ if $b \neq c$ ($b\in B$). Thus we may conclude that for every $a \in B$, \vspace{-3pt}
\begin{align*}
     g^*(a,xy,d) &= \max\{g(a,x,b) \land g(b,y,d) \mid b \in B\}\vspace{-3pt}\\
        &= g(a,x,c) \land g(c,y,d)
        = g(a,x,c) \land g(c,x,c) \land g(c,y,d)\\\vspace{-3pt}
        &= g^*(a,xxy,d) = s > 0.
\end{align*}
On the other hand, for any  $d' \neq d$, \vspace{-3pt}
\begin{align*}
     g^*(a,xy,d') &= \max\{g(a,x,b) \land g(b,y,d') \mid b \in B\} \vspace{-3pt}\\
     &= g(a,x,c) \land g(c,y,d') = g(a,x,c) \land g(c,x,c) \land g(c,y,d')\vspace{-3pt}\\
     &\leq g^*(a,xxy,d') = 0.
\end{align*}
Thus $\calG_{a,xy} = \{d/s\}$, for every $a\in B$. This would mean that $xy \in D\!D_1(\calG)$ although $xy \notin D\!D_3(\calF)$ (as $\calF_{p,xy} = \es$). 
\ep

\begin{proposition}\label{pr:LDD1 and LDD3 notsubset LDD2}
\, $\LDD_1 \nsubseteq \LDD_2$ and $\LDD_3 \nsubseteq \LDD_2$.
\end{proposition}

\bp 
The noninclusion $\LDD_1 \nsubseteq \LDD_2$ must hold because $\LDD_1 \se \LDD_2$ would imply that $n\LDD_1 = n\LDD_1 \cap \LDD_1 \se n\LDD_1\cap\LDD_2 = \DW \subset n\LDD_1$, a contradiction.

To prove $\LDD_3 \nsubseteq \LDD_2$, consider the FFA $\calF = (\{a,b\},\{x,y\},f)$, where $f$ is defined by $f(a,x,a) = f(a,x,b) = f(a,y,a) = f(b,x,b) =1$. Assume then that $DD_3(\calF) = DD_2(\calG)$ for some FFA $\calG$. Now $x \in DD_3(\calF)$ because $\calF_{a,x} = \{a/1,b/1\}$ and $\calF_{b,x} = \{b/1\}$. However, $x \in DD_2(\calG)$ would imply $xy \in DD_2(\calG)$ although $xy \notin DD_3(\calF)$.
\ep 

\begin{proposition}\label{nDD1 incomp DD1 cap DD2 cap DD3}
\,$n\LDD_1 \parallel \LDD_1 \cap \LDD_2 \cap \LDD_3$.
\end{proposition}

\bp If $n\LDD_1 \se \LDD_1 \cap \LDD_2 \cap \LDD_3$, then $n\LDD_1 \se \LDD_2$. By Proposition \ref{pr:nDD1 and DD2 equals nDD1 and DD3 equals DW}, this would imply that $\DW = n\LDD_1 \cap \LDD_2 = n\LDD_1$, which contradicts Proposition \ref{pr:DW, nDDi and DDi}\,(a).
On the other hand, if $\LDD_1 \cap \LDD_2 \cap \LDD_3 \se n\LDD_1$, then  \vspace{-3pt}
\begin{align*}
\LDD_1 \cap \LDD_2 \cap \LDD_3 &= n\LDD_1 \cap \LDD_1 \cap \LDD_2 \cap \LDD_3\vspace{-3pt}\\
&= n\LDD_1\cap  \LDD_2 \cap \LDD_3 \vspace{-3pt}\\
&= \DW \cap \LDD_3\vspace{-3pt} = \DW,
\end{align*}
by Propositions \ref{pr:DW, nDDi and DDi}\,(a) and \ref{pr:nDD1 and DD2 equals nDD1 and DD3 equals DW}. This would contradict Proposition \ref{pr:DW sub DD1 cap DD2 cap DD3}.
\ep


\section{Classes of DD$i$--directable FFAs}\label{se:Classes of DDi-dir FFAs}
In this section we shall consider the classes of FFAs and normal FFAs with the various DD$i$--directability properties. For each $i =1,2,3$, let

(1) $\DD(i)$ denote the class of all DD$i$--directable FFAs, and

(2) $\nDD(i)$ denote the class of all DD$i$--directable normal FFAs.\\
Furthermore, let $\Dir$ be the class of all directable DFAs; recall that in Section \ref{se:Families} we interpreted with each DFA $\calA$ a crisp FFA $\Afz$.

\begin{figure}[t!]
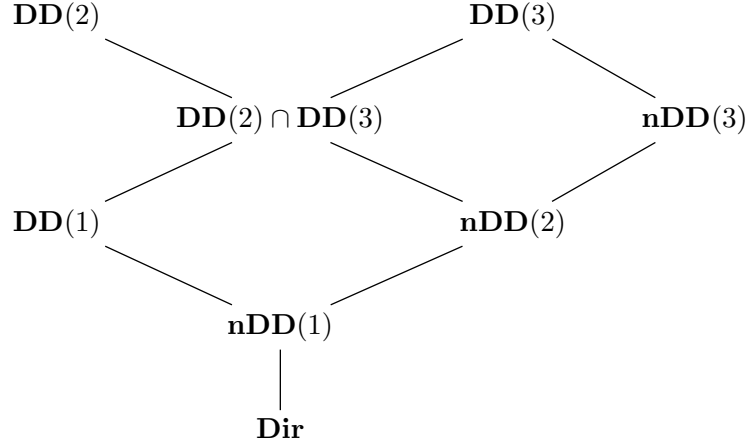


\begin{center}
$\begin{psmatrix}[colsep=10mm,rowsep=9mm]
& \rnode{DD2}{{\DD(2)}}& &\rnode{DD3}{{\DD(3)}}\\
 & & \rnode{DD2DD3}{{\DD(2)}\cap {\DD(3)}} && \rnode{nDD3}{{\nDD(3)}}\\
& \rnode{DD1}{{\DD(1)}} &&
\rnode{nDD2}{{\nDD(2)}}\\
& & \rnode{nDD1}{{\nDD(1)}}\\
& & \rnode{Dir}{\Dir}
\end{psmatrix}
$
\psset{arrows=-, nodesep=3pt,linestyle=solid,linewidth= 0.5pt, arrowscale=1.5}
\ncline{DD2}{DD2DD3}
\ncline{DD3}{DD2DD3}
\ncline{DD3}{nDD3}
\ncline{DD2DD3}{DD1}
\ncline{DD2DD3}{nDD2}
\ncline{nDD3}{nDD2}
\ncline{DD1}{nDD1}
\ncline{nDD2}{nDD1}
\ncline{nDD1}{Dir}
\end{center}

\noindent

\caption{\label{diagr1} Classes of DD-directable FFAs}
\end{figure}

The following proposition  justifies the Hasse diagram of  Figure \ref{diagr1} of the $\cap$-semilattice of the classes $\DD(i)$, $\nDD(i)$, $\Dir$ and their intersections.\smallskip

\begin{proposition}\label{pr:Classes of DDi-dir FFAs}
\begin{itemize}
  \item[{\rm (a)}] $\nDD(i) \subset \DD(i)$ for $i = 1,2,3$.
  \item[{\rm (b)}] $\DD(1) \subset \DD(2) \cap \DD(3)$.
  \item[{\rm (c)}] $\DD(2) \parallel \DD(3)$.
  \item[{\rm (d)}] $\DD(2) \cap \DD(3) \subset \DD(2)$ and $\DD(2) \cap \DD(3) \subset \DD(3)$.
  \item[{\rm (e)}] $\nDD(1) \subset \nDD(2)$.
  \item[{\rm (f)}] $\nDD(2) \subset \nDD(3)$.
  \item[{\rm (g)}] $\DD(1) \parallel \nDD(2)$.
  \item[{\rm (h)}] $\DD(2) \cap \DD(3) \parallel \nDD(3)$.
  \item[{\rm (i)}] $\nDD(2) \subset \DD(2) \cap \DD(3)$.
  \item[{\rm (j)}] $\DD(1) \cap \nDD(2) = \nDD(1)$.
  \item[{\rm (k)}] $(\DD(2) \cap \DD(3)) \cap \nDD(3) = \nDD(2)$.
  \item[{\rm (l)}] $\Dir \subset \nDD(1)$.
\end{itemize}
\end{proposition}

\bp
(a) $\nDD(i) \subset \DD(i)$ for $i = 1,2,3$:\\
The inclusions are obvious. It is easy to show that they are proper by finding for each $i = 1,2,3$ an FFA that is DD$i$--directable but not normal.\smallskip

(b) $\DD(1) \subset \DD(2) \cap \DD(3)$:\\
The inclusion follows from Proposition \ref{pr:D$i$ and DD$i$}\,(d). That it is proper, can be seen by considering the FFA $\fzF$, where $A = \{a,b\}$, $X = \{x\}$, and $f(a,x,a) = f(a,x,b) = f(b,x,a) = f(b,x,b) = 1$. For any $w = x^n$, where $n \geq 1$,  $\calF_{a,w} = \{a/1,b/1\} = \calF_{b,w}$. Thus $D\!D_2(\calF) = D\!D_3(\calF) = \{x^n \mid n \geq 1\}$, while $D\!D_1(\calF) = \es$.\smallskip

(c) $\DD(2) \parallel \DD(3)$:\\
Consider the FFA $\fzF$, where $A = \{a,b\}$, $X = \{x,y\}$, $f(a,x,a) = 0.2$, $f(a,x,b) = f(a,y,a) = f(b,x,b) = f(b,y,b) = 1$. If $w = y^n$ with $n \geq 1$, then $\calF_{a,w} = \{a/1\}$ and $\calF_{b,w} = \{b/1\}$, which means that $w$ is not DD2--directing. If the letter $x$ appears in $w \in X^*$ at least once, then $\calF_{a,w} = \{a/0.2,b/1\}$ and $\calF_{b,w} = \{b/1\}$, and thus $w \in D\!D_3(\calF)\setminus D\!D_2(\calF)$. This means that $\calF \in \DD(3) \setminus \DD(2)$.

Next, let $\calG = (\{a,b\},\{x\},g)$ be the FFA, where the only nonzero transition is $g(a,x,b) = 1$. Obviously, $\calG_{a,x} = \{b/1\}$, $\calG_{b,x} = \es$, and $\calG_{a,w} = \calG_{b,w} = \es$ for any $w = x^n$ with $n \geq 2$. This shows that $\calG \in \DD(2) \setminus \DD(3)$. \smallskip

(d) $\DD(2) \cap \DD(3) \subset \DD(2)$ and $\DD(2) \cap \DD(3) \subset \DD(3)$:\\
The properness of these inclusions follows from (c).\smallskip

(e) $\nDD(1) \subset \nDD(2)$:\\
By (b), any DD1--directable FFA is also DD2--directable, and naturally this applies to normal FFAs, too. The properness of the inclusion  is shown by the FFA $\calF$ used in the proof of (b) as it is also normal.\smallskip

(f) $\nDD(2) \subset \nDD(3)$:\\
Let $\fzF$ be a DD2--directable normal FFA, and consider any $w \in D\!D_2(\calF)$ and $b \in A$. Since $\calF$ is normal, $f^*(b,w,c) = 1$ for some $c \in A$. As $w$ is DD2--directing, this implies that $f^*(a,w,c) = 1$ for every $a \in A$, and thus $w$ is also DD3--directing. To show that the inclusion $\nDD(2) \se \nDD(3)$ is proper, consider the  FFA $\calF \in \DD(3) \setminus \DD(2)$ used in the proof of (c).  Since $\calF$ is normal, also $\calF \in \nDD(3) \setminus \nDD(2)$ holds. \smallskip

(g) $\DD(1) \parallel \nDD(2)$:\\
Consider the FFA $\fzF$, where $A = \{a,b\}$, $X = \{x,y\}$, and $f$ is given by $f(a,x,b) = f(b,x,b) = f(b,y,b) = 1$. It is in $\DD(1)$ because $\calF_{a,x} = \calF_{b,x} = \{b/1\}$, but it is not in $\nDD(2)$ as it is not normal. On the other hand, $\nDD(2) \se \DD(1)$ cannot hold either because $\nDD(2) \setminus \nDD(1) \neq \es$ by statement (e).\smallskip

(h) $\DD(2) \cap \DD(3) \parallel \nDD(3)$:\\
The FFA $\calF = (\{a,b\},\{x\},f)$ with $f$ defined by $f(a,x,b) = f(b,x,b) = 0.5$ is DD2-- and DD3--directable, but it is not normal. The FFA  $\calG = (\{a,b,c\},\{x\},g)$, in which $g$ is given by $g(a,x,b) = 0.5$ and $g(a,x,a) = g(b,x,c) = g(c,x,c) = 1$, is normal and DD3--directable. For example, $xx \in D\!D_3(\calG)$. On the other hand, there are no DD2--directing words as $\calG_{a,x} = \{a/1,b/0.5\}$, $\calG_{b,x} = \calG_{c,x} = \{c/1\}$, and $\calG_{a,w} = \{a/1,c/0.5\}$, $\calG_{b,w} = \calG_{c,w} = \{c/1\}$ for any $w = x^n$ with $n \geq 2$.\smallskip

(i)  $\nDD(2) \subset \DD(2) \cap \DD(3)$:\\
Of course, $\nDD(2) \se \DD(2)$, and by using statements (f) and (a), we get $\nDD(2) \se \nDD(3) \se \DD(3)$. Thus $\nDD(2) \se \DD(2) \cap \DD(3)$. The FFA $\calF$ used in the proof of (h) is DD2-- and DD3--directable, but it is not normal. Hence, the inclusion is proper.\smallskip

(j) $\DD(1) \cap \nDD(2) = \nDD(1)$:\\
The inclusion $\nDD(1) \se \DD(1) \cap \nDD(2)$ holds because $\nDD(1) \se \DD(1)$ by (a), and $\nDD(1) \se  \nDD(2)$ by (e). On the other hand, if $\calF \in \DD(1) \cap \nDD(2)$, then $\calF$ is both DD1--directable and normal, i.e., $\calF \in \nDD(1)$.\smallskip

(k) $(\DD(2) \cap \DD(3)) \cap \nDD(3) = \nDD(2)$:\\
Obviously, (k) is equivalent to the equality $\DD(2) \cap \nDD(3) = \nDD(2)$ which has a similar proof as (j).\smallskip

(l) $\Dir \subset \nDD(1)$:\\
Any DFA is  clearly normal when viewed as an FFA, and by Lemma \ref{le:DFA to FFA} is DD1--directable. On the other hand, the FFA $\fzF$, where $A = \{a,b,c\}$, $X = \{x\}$, and $f$ is given by $f(a,x,b) = f(a,x,c) = f(b,x,c) = f(c,x,c) = 1$, is DD1--directable and normal, but not a DFA.
\ep


Let us now consider the subautomata, epimorphic images and finite direct products of DD1--, DD2-- and DD3--directable FFAs. In what follows, $\fzF$ and $\fzG$ are any two FFAs.

We call $\calG$ a \emph{subautomaton} of $\calF$ if (1) $B \se A$, (2) $\calF(b,x) \se B$ for all $b\in B$ and $x \in X$, and (3) $g(b,x,b') = f(b,x,b')$ for all $b,b' \in B$ and $x \in X$.

\begin{lemma}\label{le:Subautomaton}
Let $\calG$ be a subautomaton of  $\calF$. For any $b\in B$ and $w \in X^*$, \vspace{-3pt}
\begin{itemize}
  \item[{\rm (a)}] $\calF(b,w) \se B$,\vspace{-3pt}
  \item[{\rm (b)}] $\calG(b,w) = \calF(b,w)$, and \vspace{-3pt}
  \item[{\rm (c)}] $\calG_{b,w} = \calF_{b,w}|_B$ (= the restriction of $\calF_{b,w}$ to $B$).\vspace{-3pt}
\end{itemize}
\end{lemma}

\bp All three statements have easy proofs by induction on $\lg(w)$. \ep



\begin{proposition}\label{pr:DDi and subautomata}
Let $\calG$ be a subautomaton of an FFA $\calF$. For each $i = 1,2,3$, $D\!D_i(\calF) \se D\!D_i(\calG)$, and thus any subautomaton of a DD$i$--directable FFA is also DD$i$--directable. Furthermore, if $\calF$ is normal, then so is $\calG$.
\end{proposition}

\bp  
If $w \in D\!D_1(\calF)$, then there exist $c \in A$ and $r \in (0,1]$ such that $\calF_{a,w} = \{c/r\}$ for every $a \in A$. In particular, for any $b \in B$, $c \in \calF(b,w) \se B$, and hence $\calG_{b,w} = \calF_{b,w}|_B = \{c/r\}$. This means that $w \in D\!D_1(\calG)$.

If $w \in D\!D_2(\calF)$, then $\calG_{a,w} = \calF_{a,w}|_B = \calF_{b,w}|_B = \calG_{b,w}$ for all $a,b\in B$, i.e., $w \in D\!D_2(\calG)$.

Let $w \in D\!D_3(\calF)$, and let $c \in A$ be such that for every $a \in A$, $\calF_{a,w}(c) >0$ and $\calF_{a,w}$ is maximal among all the values of $\calF_{a,w}$. Since this holds also for every $a \in B$, we have $c \in B$, and therefore $w \in D\!D_3(\calG)$.

Assume now that $\calF$ is normal, and consider any $b\in B$ and $x \in X$. Since $\calF$ is normal, there is a state $c \in A$ such that $f(b,x,c) = 1$. By Lemma \ref{le:Subautomaton}, $c \in B$ and $g(b,x,c) = f(b,x,c) = 1$, which shows that $\calG$ is normal.
\ep
\vspace{3pt}

A mapping $\vp : A \ra B$ defines a \emph{homomorphism} $\vp : \calF \ra \calG$ of FFAs (cf. \cite{MaMS97,MoMa02,Pet06}, for example) if \vspace{-3pt}
\[
g(a\vp,x,b) = \max\{f(a,x,a') \mid a' \in b\vpi\}, \vspace{-3pt}
\]
for all $a \in A$, $b \in B$ and $x \in X$. A surjective homomorphism is called an \emph{epimorphism}.

\begin{lemma}\label{le:Homomorphisms of FFAs} If $\vp : \calF \ra \calG$ is an epimorphism of FFAs, then  for all $a \in A$, $b \in B$ and $w \in X^*$, \vspace{-3pt}
\begin{itemize}
  \item[{\rm (a)}] $g^*(a\vp,w,b) = \max\{f^*(a,w,a') \mid a' \in b\vpi\}$, and \vspace{-3pt}
  \item[{\rm (b)}] $\calF(a,w)\vp = \calG(a\vp,w)$.\vspace{-3pt}
\end{itemize}
\end{lemma}

\bp Statement (a) has a straightforward proof by induction on $\lg(w)$.

Let us consider (b). Also here we proceed by induction on $\lg(w)$.

If $a' \in \calF(a,w)$, then\vspace{-3pt}
\[
g^*(a\vp,w,a'\vp) = \max \{f^*(a,w,a'') \mid a''\in A, a''\vp = a'\vp\} \geq f(a,w,a') > 0 \vspace{-3pt}
\]
implies that $a'\vp \in \calG(a\vp,w)$, and hence that $\calF(a,w)\vp \se \calG(a\vp,w)$. Conversely, if $b \in \calG(a\vp,w)$, then\vspace{-3pt}
\[
\max \{f^*(a,w,a') \mid a' \in b\vpi\} = g^*(a\vp,w,b) > 0 \vspace{-3pt}
\]
means that $f^*(a,w,a') >0$, i.e., $a' \in \calF(a,w)$, for some $a' \in b\vpi$, and thus $b \in \calF(a,w)\vp$.
\ep

\begin{proposition}\label{pr:Morphims and DDi}
If $\vp : \calF \ra \calG$ is an epimorphism of FFAs, then $D\!D_i(\calF) \se D\!D_i(\calG)$ for $i = 1,2,3$, and thus all epimorphic images of a DD$i$--directable FFA are DD$i$--directable. Moreover, if $\calF$ is normal, then so is $\calG$.
\end{proposition}

\bp  Let us prove $D\!D_i(\calF) \se D\!D_i(\calG)$ for each value of $i$.

If $w \in D\!D_1(\calF)$, then there exist a $c \in A$ and an $r \in (0,1]$ satisfying $\calF_{a,w} = \{c/r\}$ for every $a \in A$. Consider any $b \in B$. Since $\vp$ is surjective, $b = a\vp$ for some $a \in A$. For any $d \in B$, \vspace{-3pt}
\[
g^*(b,w,d) = \max\{f^*(a,w,a') \mid a' \in d\vpi\} = \left\{\begin{array}{cl}
                         r,&\ \mbox{if } d = c\vp\\
                         0,&\  \mbox{if } d \neq c\vp. \vspace{-3pt}
                         \end{array}
                   \right.
\]
This means that $\calG_{b,w} = \{c\vp/r\}$ for every $b \in B$, and thus $w \in D\!D_1(\calG)$.

If $w \in D\!D_2(\calF)$, then $f^*(a_1,w,a) = f^*(a_2,w,a)$ for all $a_1,a_2,a \in A$. Consider any states $b_1,b_2 \in B$ of $\calG$, and let $a_1,a_2 \in A$ be states of $\calF$ for which $b_1 = a_1\vp$ and $b_2 = a_2\vp$. For any $b\in B$,
\begin{align*}
     g^*(b_1,w,b) &=  \max\{f^*(a_1,w,a) \mid a \in b\vpi\} \vspace{-3pt}\\
        &= \max \{f^*(a_2,w,a) \mid a\in b\vpi\} =  g^*(b_2,w,b), \vspace{-3pt}
\end{align*}
which shows that $w \in D\!D_2(\calG)$.

Let $w \in D\!D_3(\calF)$. Then there is a $c \in A$ such that, for every $a \in A$, $f^*(a,w,c) > 0$ and $f^*(a,w,d) \leq f^*(a,w,c)$ for every $d \in A$. Consider any $b,b'\in B$, and let $a,a'\in A$ be states for which $b = a\vp$ and $b' = a'\vp$. Then \vspace{-3pt}
\[
g^*(b,w,c\vp) = \max\{f^*(a,w,a'') \mid a''\vp = c\vp\} \geq f^*(a,w,c) >0, \vspace{-3pt}
\]
and
\[
g^*(b,w,b') = \max\{f^*(a,w,a'') \mid a'' \in b'\vpi\} \leq f^*(a,w,c) \leq g^*(b,w,c\vp),
\]
which means that $w \in D\!D_3(\calG)$.

Assume now that $\calF$ is normal, and consider any $b\in B$ and $x \in X$. If $a \in A$ and  $c \in A$ are states for which $a\vp = b$ and  $f(a,x,c) = 1$, then
\[
g(b,x,c\vp) = g(a\vp,x,c\vp) = \max\{f(a,x,d) \mid d\vp = c\vp\} \geq f(a,x,c) = 1,
\]
i.e., $g(b,x,c\vp) = 1$.
\ep

The \emph{direct product} (cf. \cite{MaMS97,MoMa02}, for example) of two FFAs $\calF$ and $\calG$ is the FFA $\calF \times \calG = (A \times B,X,h)$, where $h$ is defined by \vspace{-3pt}
\[
h((a,b),x,(a',b')) = f(a,x,a') \land g(b,x,b') \quad (a,a'\in A, b,b'\in B, x\in X). \vspace{-3pt}
\]

The following example shows that DD$i$--directability ($i = 1,2,3$) is not preserved under this product.

\begin{example}\label{ex:DDi and direct products}{\rm
Let $\calF = (\{a,b\},\{x,y\},f)$ and $\calG = (\{1,2\},\{x,y\},g)$ be FFAs, where $f$ and $g$  are given by $f(a,x,b) = f(b,x,b) = f(b,y,b) = 1$, and $g(1,y,2) = g(2,y,2) = g(2,x,2) = 1$. These FFAs belong to all three classes $\DD(1)$, $\DD(2)$ and $\DD(3)$ because $x \in D\!D_1(\calF) \cap D\!D_2(\calF) \cap D\!D_3(\calF)$ and $y \in D\!D_1(\calG) \cap D\!D_2(\calG) \cap D\!D_3(\calG)$.

As $(\calF \times \calG)((a,1),x) = (\calF \times \calG)((a,1),y) = \es$ while $(\calF \times \calG)((b,2),x) = (\calF \times \calG)((b,2),y) = \{(b,2)\} \neq \es$, it is clear that $\calF \times \calG$ has no DD1--, DD2-- or DD3--directing words, and
thus $\calF \times \calG \notin \DD(1) \cup \DD(2) \cup \DD(3)$.
}
\end{example}


\end{document}